\documentstyle[aps]{revtex}
\input epsf

\begin{document}
\twocolumn[
\preprint{FERMILAB-Pub-96/054-E \& WM-99-102}

\title {\bf Experimental and Theoretical Results for Weak Charge Current
Backward Proton Production}

\author{C. E. Carlson}
\address{Physics Department, College of William and Mary,
Williamsburg, VA 23187, USA}

\author{J. Hanlon}
\address{Fermi National Accelerator Laboratory, Batavia,  IL,  60510}

\author{K. E. Lassila}
\address{Department of Physics and Astronomy, Iowa State University,
Ames, IA 50011}

\date{February 1999}

\maketitle
 
\vskip -0.2in
 
\begin{abstract} \widetext
 
\parshape=1 0.5in 6in
In this paper, we do three things in the study of deuteron break-up by
high energy neutrino beams.  (1) We present previously unpublished
data on neutrino induced backward protons from deuteron targets; (2) we
calculate the contributions from both the two-nucleon (2N) and
six-quark (6q) deuteron components, which depend upon the overall
normalization of the part that is 6q; and (3) we suggest other signatures
for distinguishing the 2N and 6q clusters.  We conclude that the 6q
cluster easily explains the shape of the high momentum backward proton
spectrum, and its size is nicely  explained if the amount of 6q is one
or a few  percent by normalization of the deuteron.  There is a
crossover, above which the 6q contribution is important or dominant, at
300--400 MeV/c backward proton momentum.

\end{abstract}
 
\vskip 0.2in
 

\vskip -35pt

]

\narrowtext

\section {Introduction}
  
        It is worthwhile to examine and probe the deuteron as a target from
many viewpoints because of the role it plays as our main source of
information on the neutron.  This has become of increasing interest
in recent years with unexpected results being found for the
numerous sum rules in which neutron structure functions enter. 
These include the results for the Ellis-Jaffe sum
rules~\cite{ash89,bau83}, for the Gottfried sum rule~\cite{ama91},
and for the Bjorken sum rule~\cite{ade93,ant93},  prompting us to
re-examine assumptions made in most analyses of deuteron data to
extract neutron information.  The deuteron is more than the sum of
its proton and neutron parts and since experiments on free neutrons
can not be done, this nonadditive portion is important to evaluate
by any means available.  We believe, in fact, that earlier
experiments~\cite{aub83,arn84} make it clear that the deuteron is
the simplest nucleus exhibiting the EMC effect which, in turn,  must
affect the extraction of the neutron structure function that enters
in these sum rules~\cite{las91}.

In the theoretical part of the present paper, we enlarge upon the work
we introduced earlier~\cite{car91} and apply it to some old data
and to new, previously unpublished data.  We have studied deep
inelastic $\nu$ and $\bar \nu$ scattering from various targets,
focusing on reactions that produce high momentum backward protons. 
Backward means relative to the incoming neutrino or antineutrino and
high momentum means relative to kinematic limits upon backward
momentum imposed in terms of quantities such as the target mass and
momentum fraction carried by the struck quark.  

We concluded, based on deuteron target data~\cite{mat89} obtained at
CERN with $\nu$ and $\bar \nu$ beams, that the momentum distribution
of the backward protons was consistent with production from a
multiquark component (6q) in the deuteron and was difficult to explain
if produced via break-up of a two nucleon (2N) quantum mechanical state
with a simple conventional wave function.  The particular problem
with the latter picture, where the neutron and proton
substantially maintain their character as nucleons~\cite{fra77}, was
that there were more large momentum backward protons than expected
from typical neutron-proton wave functions.  However, in our
previous work we were only able to calculate the shape of the
backward proton spectrum in the two cases and not the absolute
normalization or even the relative normalization of the 6q and 2N
contributions.

In this paper, we will present normalized calculations
for backward proton production in deep inelastic experiments for
both the 2N and 6q deuteron components.  The 6q calculation includes
factors of the fraction of the deuteron that is 6q and of the
fragmentation rate of the residue of the 6q cluster into protons.  A
numerical uncertainty in the latter rate for high momentum protons
leads to a factor circa 2 uncertainty in fixing the fraction of
6q state in the deuteron;  otherwise the calculation is well
determined.

We denote the probability of finding the 6q configuration in the
deuteron as $f$, with $f$  expected to be between 0.01 and 0.07. 
Possible values for $f$  have been calculated from deuteron wave
functions as the probability for the nucleons to overlap, and Sato et
al.~\cite{sat86} claim a value for $f$  in the middle of this range
for a typical nucleon-nucleon potential.  The largest deuteron
probability, $f = 0.074$, was used~\cite{yen89} in describing high
energy SLAC electron-deuteron data at $x > 1.0$.  Also, studies of
the deuteron electromagnetic structure functions have been used to
estimate $f$ to be a few percent~\cite{kobushkin}. We here will find
that the 6q contribution in deep inelastic scattering, even with $f$
only one or two percent, can be quite large for energetic backward
protons. Throughout, although we are mindful of many possibilities
(see e.g.,~\cite{manton}), we simplify by speaking of the short
range baryon number correlation as either 6q or 2N.  Within this
limitation, we can further say that for most of the conventional 2N
wave functions, the 6q state not only can but also must contribute
a major share of the cross section for energetic backward protons.
We will show that at 500 MeV/c backward proton momentum the modern
nucleon-nucleon potential that comes closest to our data needs at least 
a 60\% additional contribution.

     On the experimental side, we shall in the next Section discuss a
Fermilab neutrino-deuterium 15-ft bubble chamber experiment (E545)
which has obtained data (previously unpublished) on backward
production of protons.  Also,  we shall comment
on some published experiments which likewise measured high-momentum
backward proton events.  

In Sect.~\ref{three}, we give the theoretical formulae used in
calculation of backward proton spectra to compare with the Fermilab
and CERN deuteron break-up cross sections for backward protons.  A
comparison of the new data with previously published Argonne,
Brookhaven, CERN, and Fermilab backward proton production data is
also given.  The comparisons of the prediction given by modern
potentials with our data is in Sect.~\ref{four}.  The final Section is 
devoted to further discussion and
presentation of conclusions.

\section{Fermilab E545 neutrino-deuterium data} \label{two}

        Previous analyses of the E545 neutrino-deuterium data have
concentrated on either extracting $\nu n$ interaction data, or on the
distributions of lepton or hadron variables in $\nu d$ scattering. A
typical analysis would separate the data into even-prong ``$\nu n$''
and odd-prong ``$\nu p$'' events, where a visible proton spectator is
ignored in the prong count. The observed ``$\nu n$'' events were
assumed to be a sample of $\nu n$ interactions depleted by
``rescattering'' within the deuteron nucleus. The ``rescattered''
$\nu n$ events, in turn, would appear in the ``$\nu p$'' event sample.
In extraction of ratios of $\nu n$ to $\nu p$ cross sections the
fraction of such rescattered events was estimated to be in the
6--12\% range.

An excess of high momentum proton spectators, compared to standard
deuteron wave function predictions, is known to be associated with
the ``$\nu n$'' event sample. When commented on in previous analyses,
it would generally be noted that the origin of this excess is
unknown, but given that the excess accounts for less than 1\% of the
``$\nu n$'' events any effect on the overall distributions or measured
quantities would be negligible. In the present analysis, we
explicitly examine the $\nu d$ target fragments in the E545 data,
and compare the distribution of backward protons with a
deuteron wave function plus a small six-quark component. 

Many details of the E545 experiment have been
published~\cite{cha83}. The data discussed here were previously
presented at an APS meeting~\cite{kaf83}, where the
emphasis was on establishing the existence of a ``rescattering''
phenomena which depleted the observed spectator proton (neutron
target) event sample, and in estimating its frequency.

The E545 data are from a 320,000 frame exposure of the
deuterium-filled Fermilab 15-ft bubble chamber to a wide-band
single-horn focused neutrino beam produced by 
$4.8 \times 10^{18}\ 350$ GeV/c protons incident on a beryllium
oxide target. The anti-neutrino component of the beam is $\approx
14$\%. The film was scanned twice, and events with two or more
charged tracks produced by incident neutral particles in a 15.6
m$^3$ fiducial volume were accepted for analysis. All charged
tracks were digitized and geometrically reconstructed.
Topology-dependent weights are applied to the data to compensate
for scanning and processing losses and for those events failing
geometric reconstruction. The average processing times scanning
efficiency is 0.80. Cuts are applied to the two-prong events to
remove $K^0$ and
$\Lambda$ decays and $\gamma$ conversions from the data. 

A kinematic technique which uses only the measured momenta of the
charged particles is used to select a sample of charge current
events. Only events for which $\sum p_L > 5$ GeV/c, where $p_L$ is
the component of laboratory momentum in the beam direction and the
sum is taken over all charged particles, are included in the
analysis. The muon candidate is identified as that negative track
in the event with the largest component of momentum transverse to
the incident neutrino direction. Those events for which the
component of the $\mu^-$ candidate's momentum transverse to the
vector sum of the momenta of the other charged particles in the
event is greater than 1.0~GeV/c are accepted as charge current
events. 

The incident neutrino energy of the selected charge current
events is estimated using transverse momentum balance: 
$E_\nu = p_L^\mu + p_L^H +|\vec p_T^{\,\mu} +\vec p_T^H| p_L^H/p_T^H$,
where the symbols $p^\mu$ and $p^H$ refer to the muon momentum and the
vector sum of the charged hadron momenta, respectively. Only events
with $E_\nu > 10$ GeV are accepted for analysis.  

A Monte Carlo simulation indicates that the sample selected
according to the above criteria includes 79\% of the $\nu d$
charge current events, with the $\mu^-$ correctly identified in 98\%
of the cases, and with a 3\% contamination due to $\nu d$
neutral-current events and 1\% due to $\bar\nu d$ events.

The corrected number of $\nu d$ events in the sample is 15,129, with
an average neutrino energy $\langle E_\nu \rangle = 50$ GeV. Of these
events, 459 have an identified proton with momentum magnitude greater
than 160 MeV/c whose direction is backward with respect to the
incident neutrino direction. (Significant visibility losses occur for
protons with momentum less than 160 MeV/c, and hence are not
presented.) The identity of the backward protons was verified by
re-examining all such tracks on the scan table. The momentum
distribution of the backward protons is given in Table~\ref{table}.
This data will be discussed in Sect. 3, together with the proton
spectrum from a $\nu d$ and $\bar\nu d$ exposure of BEBC by the WA25
collaboration at CERN~\cite{mat89}.

        The published Fermilab E545 spectator proton spectra from
quasi-elastic $\nu d$ scattering 
($\nu d \rightarrow \mu^- p p_s$)~\cite{kit83} will also be
discussed, together with similar distributions from $\nu d$ bubble
chamber experiments at Brookhaven (80-in)~\cite{baker} and Argonne
(12-ft)~\cite{bar77}.

\section{Theoretical Discussion}             \label{three}

     We will give the expressions for the charge current inclusive cross
sections of neutrinos hitting a deuteron and producing a
backward proton, $p_B$, a forward lepton, $\ell^-$, and anything 
else, $X$,
\begin{equation}
\nu + d \rightarrow \ell^- + p_B + X,
\end{equation}

\noindent in both the 2N and 6q models.   We
will see that the shape of the backward proton spectrum is
different in the two cases, and will see if the calculations can
match the data.  The two contributions are not mutually exclusive,
and we shall add them incoherently, weighting the 6q
contribution by fraction $f$ and the 2N contribution by $(1 - f)$. 
We expect that $f$ will be on the order of a few percent,  but that
none-the-less the 6q contribution could be large at large backward
proton momenta.

The fully differential cross section is differential
in $x$, $y$, $\alpha$, and $p_T$,  which are the
experimentally measurable variables.  These variables are the
struck quark momentum fraction,
\begin{equation}
x = Q^2/2m_N \nu = Q^2/2m_N (E_\nu - E_\ell);
\end{equation}

\noindent the fractional lepton energy loss,
\begin{equation}
y  = {E_\nu - E_\ell \over E_\nu };
\end{equation}

\noindent the light front momentum fraction of the backward proton,
\begin{equation}
0 \leq \alpha = {E_p + p_z \over m_N } \leq 2
\end{equation}

\noindent (with $p_z$ defined positive for backward protons);  and
the transverse momentum of the proton relative to the
direction of the incident neutrino, $p_T$.

For the 2N model, using quark distribution functions appropriate to
describe  striking the neutron, with the argument changed from $x$
to
$\xi= x/(2 - \alpha)$ because the neutron is moving, and having
the proton emerge with probability given in terms of the deuteron
wave function, we find the cross section

\begin{eqnarray}
{{d\sigma _{2N}} 
         \over {dx\,dy\,d\alpha\,d^2p_T}}
  &=&
        \sigma_0  \times  \\ \nonumber
        &\times&
        \left( {D_n(\xi)+S_n(\xi)+(1-y)^2 \bar U_n(\xi)} \right)
          \\ \nonumber
        &\times&
         {(2-\alpha ) \over \gamma}
        \left| {\psi (\alpha,p_T)} \right|^2 ,
\end{eqnarray}

\noindent where $D_n(\xi)$ is $\xi$ times the distribution
function of down quarks in the neutron ($n$),  etc.,
$\gamma = E_p/m_N$, $\sigma_o$ is the point Fermi weak interaction (with
strenght $G_F$) cross section, 
\begin{equation}
\sigma_0 \equiv {{2G_F^2 m_N E_\nu}\over \pi } ,
\end{equation}
  
\noindent and $\psi$ is the wave function of the deuteron
normalized by
\begin{equation}
\int d\alpha \, d^2 p_T \ \left| {\psi (\alpha,p_T)} \right|^2 =1.
\end{equation}

The corresponding cross section for the 6q component
of the target can be written in terms of the probability
distribution of a quark in the 6q cluster and in terms of the
probability, $D_{p/5q}$, for the residuum of the 6q
state to fragment into the proton.  This time, since the deuteron
or 6q cluster is stationary in the lab, $x$ is directly---in the
scaling limit---the momentum fraction of the struck quark.  We have
\begin{equation}
{{d\sigma _{6q}} \over {dx\,dy\,d\alpha
     \,d^2p_T}} = \sigma_0 D_6(x) \cdot 
    {1 \over {2-x}} D_{p/5q}(z,p_T)  ,
\end{equation}

\noindent where we have included just $D_6$,  the down quark
distribution for the 6q cluster times $x$,  on the grounds that we
will need the literal 5q residuum (which comes from the 6q Fock
component of the nominal 6q cluster) to get the highest momenta
backward protons.  The first argument of the fragmentation
function is the light front momentum fraction of the proton relative
to the five quark residuum, or
\begin{equation}
z = {\alpha \over 2-x}.
\end{equation}

\begin{table} 
\begin{center}
\begin{tabular}{cc}
Momentum range (MeV/c) & Number of events \\
160-200  &  187  \\
200-240  &  100  \\
240-280  &   61  \\
280-320  &   37  \\
320-360  &   30  \\
360-400  &   14  \\
400-440  &   10  \\
440-480  &   10  \\
480-520  &    9  \\
520-560  &    1  \\
560-600  &    0  \\
\end{tabular}

\end{center}

\caption{Momentum distribution of backward protons in 15,129 $\nu d$
charge current events from Fermilab experiment E545.}
\label{table}
\end{table}

Presently reported data on the backward proton momentum spectrum
uses protons gathered from the entire backward hemisphere.  Hence, we
too will integrate over the backward hemisphere, to allow direct
comparison to experiment.  We will integrate over $x$ and $y$ also. 
The term with explicit $y$ dependence gives a small contribution.  
Then,
\begin{eqnarray}
{{E_p \over  p^2} {d\sigma_{2N}\over dp}} &=& 
   \int_{bkwd} d\Omega\, dx\  E_p {{d\sigma _{2N}} \over {d^3p} \, dx}
                            \nonumber \\
&=&   \sigma_0 {\bar \xi_{\nu n}} \int_{bkwd} d\Omega \, dx
\ \gamma^{-1}\alpha (2-\alpha )^2
\left| {\,\psi \,} \right|^2
\end{eqnarray}

\noindent where
\begin{equation}
\bar \xi_{\nu n}  =  \int_0^1  d\xi \,
\left(D_n(\xi)+S_n(\xi)+{1\over 3} \bar U_n(\xi) \right)
\end{equation}

\noindent and

\begin{eqnarray}
{{E_p \over  p^2} {d\sigma_{6q}\over dp}} &=&
   \int_{bkwd} d\Omega\, dx \ E_p{{d\sigma _{6q}} \over {d^3p} \, dx}
                               \nonumber \\
&=&  \int_{bkwd}  d\Omega \, dx \ \alpha D_6(x) D_{p/5q}(z,p_T),
\end{eqnarray}

\noindent where $p = |\vec p \,|$ is the backward proton momentum
(and we used 
$\alpha d\sigma /d\alpha\,d^2p_T = Ed \sigma /d^3p$).  
What we want is the weighted sum of the 2N and 6q contributions,
\begin{equation}
{d\sigma \over dp} = 
      (1-f) {d\sigma_{2N} \over dp}
      + f   {d\sigma_{6q} \over dp}  .
\end{equation}

The plotted curves are based on the above formulas
plus some choices for the deuteron wave function, the quark
distribution functions, and the fragmentation function of the 5q
residuum.  

The wave function will be a light front wave function.  (A simple use
of a non-relativistic wave function conflicts with the kinematic
bound on maximum backward proton momentum.)  It is related to
non-relativistic wave functions by
\begin{equation}
\left| \psi(\alpha, p_T) \right|^2 = 
      \left| \psi_{LF}(\alpha, p_T) \right|^2
         = {E_k\over \alpha (2-\alpha)} 
         \left| \psi_{NR}(k_z, k_T) \right|^2
\end{equation}

\noindent where the arguments of the non-relativistic wave
function are obtained from
\begin{equation}
k_T = p_T
\end{equation}

\noindent and
\begin{equation}
\alpha = {\sqrt{m_N^2 + \vec k^2}+ k_z \over 
         \sqrt{m_N^2 + \vec k^2}  }.
\end{equation}

\noindent The normalization is
\begin{equation}
\int d^3k\,  \left| \psi_{NR}(k_z, k_T) \right|^2 =1
\end{equation}

\noindent and the factor above comes from the Jacobian in
\begin{equation}
d^3k = {E_k\over \alpha (2-\alpha)} d\alpha \, d^2p_T .
\end{equation}

\noindent We use several different deuteron wave functions, but
start with a Hulth\'en wave function, which is still in common
use~\cite{tenner,kit83,baker},
\begin{equation}
\psi_{NR}(\vec k) \propto {1\over {\vec k}^2 + (45.6 {\rm\ MeV})^2}
                         -{1\over {\vec k}^2 + (260 {\rm\ MeV})^2}.
\end{equation}

The quark distributions for the nucleon are the set
CTEQ1L~\cite{cteq93}.  (Some old and simple quark
distributions~\cite{ch83} give results about the same.)  For the 6q
cluster, we use the Lassila-Sukhatme model ``B'' quark
distributions~\cite{ls88}.  These distributions are based on quark
counting rules and physical logic and describe the EMC data.  Models
``A'' and ``C'' are not very different for the present purposes and
are omitted from the figures mainly to avoid clutter.  Model ``B'' has
\begin{equation}
D_6(x) = 3 \times 1.85 \sqrt{x\over 2} 
      \left( 1 - {x\over 2} \right)^{10}.
\end{equation}

The fragmentation function for the 5q residuum is taken in a
factorized form,
\begin{equation}
D_{p/5q}(z,p_T) = { {(N+4)! \over N! \, 3!}}   \ 
      z^N (1-z)^3    
      \cdot  {2\over \pi \lambda^2}
      \left( 1+{p_T^2 \over \lambda^2} \right)^{-3}  ,
\end{equation}

\noindent where $\lambda = 310$ MeV.  
The spectrum of protons for $z \rightarrow 1$ is given by the
counting rules.  For this limit, the two quarks not in the proton
must give their momentum to the three that are, and this leads to
the factor $(1-z)^3$.  Then, barring effects external to the $5q$
residuum, the proton should have 3/5 of the residuum's momentum and
this requires $N=5$.  There is, however, some pull from the struck
quark which could increase the probability of protons going in the
forward direction.  This can be accommodated in the above
fragmentation function by reducing $N$, and we shall quote results
for both $N=3$ and $N=5$.  Lower values of $N$ increase the
cited values of $f$.

\begin{figure}

\vglue 0.1in

\epsfxsize 3.45in \epsfbox{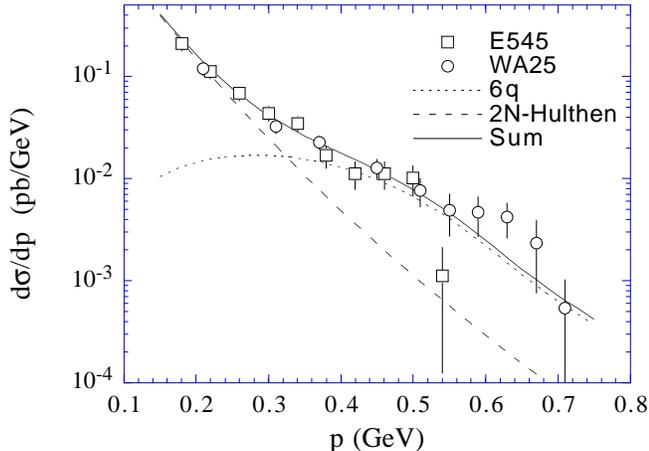}

\caption{Protons in backward hemisphere from neutrino induced
deuteron breakup; $p$ is the magnitude of the three momentum.  The
data is Fermilab 15-ft bubble chamber (E545) data and BEBC
(WA25) data.  The 6q curve as shown uses a fragmentation function
whose transverse part has a power-law falloff, 
$(1+ p_T^2/\lambda^2)^{-3}$, and whose longitudinal part goes as 
$z^5(1-z)^3$, and is normalized to give one proton per 6q cluster
breakup.  The normalization requires a 2\% fraction of 6q state. 
Using instead $z^3(1-z)^3$ leads to a curve whose shape is
essentially the same as the one shown in the important region above
250 MeV/c, and with the same normalization if the 6q fraction is
4\%. The 2N calculation uses a Hulth\'en wave
function~\protect\cite{tenner}, adjusted for use in a light front
version of a relativistic calculation. }

\label{han53}

\end{figure}

Fig.~\ref{han53} shows the comparison of the E545 data and WA25
data~\cite{mat89} with the sum of the 6q and 2N contributions, using
$N=5$ and $f = 2\%$, or equivalently for $N=3$ and $f = 4\%$.  The 
E545 data is absolutely normalized,
$d\sigma / dp = (\sigma_{tot}^{CC}(\nu d) / N_{tot}) N_{bin}/ \Delta
p$ where
$\Delta p$ is the bin width and $\sigma_{tot}^{CC}(\nu d)$ is
obtained from~\cite{kit82a} and evaluated at the 50 GeV average
$E_\nu$ in the E545 experiment.  The BEBC data WA25 is scaled to agree
with the E545 data at the lower momenta.

The match between the calculation and the
data is quite good.  A 2N contribution alone, with this wave
function,  could not match the data.  The 6q contribution, though it
is overall only 2--4\% of the normalization of the deuteron state,
contributes the major share of the cross section for energetic
backward protons.  The crossover momentum is about 300 MeV, and above
this momentum a larger and larger majority of the protons come from
the 6q cluster.  to further elaborate this point, we note that at 500
MeV ($c=1$) backward proton momentum, the Hulth\'en contribution needs 
800\% additional contribution to be in agreement with the new data.

Backward proton data from neutrino scattering is also available from
Argonne~\cite{bar77}, Brookhaven~\cite{baker}, and again Fermilab
E545~\cite{kit83} for the ``quasi-elastic'' reaction,

\begin{equation}
     \nu + d \rightarrow  \mu^-  + p + p_s  ,
\end{equation}

\noindent where $p_s$ is a label for the ``slow
protons'' or ``spectators.''  
 
\begin{figure}

\vglue .1in

\epsfxsize 3.45in \epsfbox{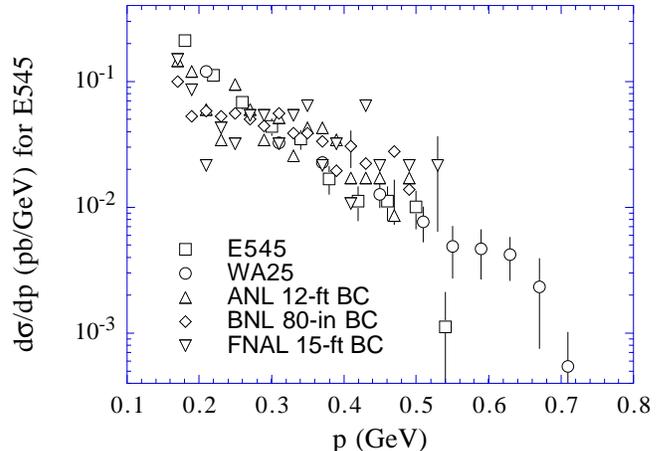}
\caption{Protons in the backward hemisphere from several different
neutrino induced deuteron breakup experiments; $p$ is the magnitude
of the three momentum.  The data labeled E545 and
WA25 are for inclusive deuteron breakup, and the data labeled ANL,
BNL, and FNAL (the latter actually also from E545) are for
quasi-elastic deuteron scattering.  The vertical scale is for the
E545 inclusive data; the other data are scaled to it.}
\label{all53}
\end{figure}

Fig.~\ref{all53} shows the backward proton spectra from the three sets
of ``quasi-elastic'' data, together with the spectra from the
inelastic data of E545 and the WA25 data.  There is reasonable
consistency within errors among all the data sets for
the proton spectrum,  despite great differences in the incoming
neutrino energy.  One may think that the material struck by the
incoming probe goes forward and that any backwardly emerging
hadrons have spectra governed only by the distribution of
constituents in the target.  The consistency among the data sets
for backward protons supports this view.  

To repeat the main point of this section, a calculation modeling
the deuteron as 2N plus a small amount of 6q is able
to match the data out to about 720 MeV backward hemisphere proton
momentum, again within experimental uncertainty.

\section{Use of other wave functions} \label{four}

The two-nucleon wave function that we have used in the above
discussion is fairly simple, and one may inquire what happens if a
more realistic wave function is used.  There are many wave
functions derived from nucleon-nucleon potentials that are fit to
nucleon-nucleon scattering data, and sometimes also to
electron-deuteron scattering data.  The assumption is made that only
nucleon-nucleon, or sometimes only baryon-baryon, degrees of
freedom are needed.  If there are other degrees of freedom
present---and that is a crucial question we are trying to
address---they are ignored.  But if they exist, their effects are
present in nature, and they must be included in the
fitted baryon-baryon potential.  And if one calculates
deuteron structure from such a potential, who is to be sure if
small effects in the (high momentum) tail of the wave function are
really due to the two nucleons, or due to the fitted wave function
trying to emulate another degree of freedom? (We should quote the
Paris group's remark that ``there is no compelling theoretical reason
to believe the validity of our potential in the region $r \leq 0.8$
fm. since the short range (SR) part of the interaction is related to
exchange of heavier systems and/or to effects of subhadronic
constituents such as quarks, gluons, etc.''~\cite{paris80}  There 
apparently would be no physical significance in the invention of a 
nucleon-nucleon potential model giving enhanced high momentum 
components by modifing this $r \leq 0.8$ fm. region.  But, the important
point is: It is precisely
this region that this our new data probes.)

\begin{figure}

\vglue .15in

\epsfxsize 3.3in \epsfbox{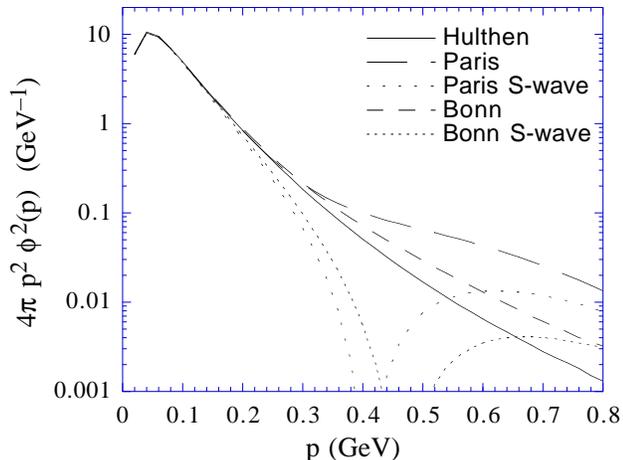}
\caption{Comparison of the Bonn, Hulth\'en, and Paris wave functions. 
(The $\phi^2$ on the vertical axis is a sum of $\phi_S^2$ and
$\phi_D^2$ for the Paris and Bonn wave functions.)}
\label{compare}
\end{figure}

Thus, if one speaks of more realistic wave functions in the present
context, one may object to the phrase ``more realistic.''  We shall
however take the Paris and Bonn wave functions~\cite{paris81,bonn87}
as representative of more sophisticated wave functions and see what
happens when we use them.  A first plot, Fig.~\ref{compare}, shows 
the Paris, Bonn, and Hulth\'en wave function in momentum space.  They
are essentially the same at low momenta, but at several hundred MeV,
thanks mainly to the shape and 5.77\% size of the D-state, the Paris
wave function is considerably larger.  Among other wave functions,
both the Reid wave function~\cite{reid} and the relatively new wave
function of Van Orden, Devine, and Gross~\cite{cebaf95} are
rather close to the Paris wave function.

\begin{figure}

\epsfxsize 3.45in \epsfbox{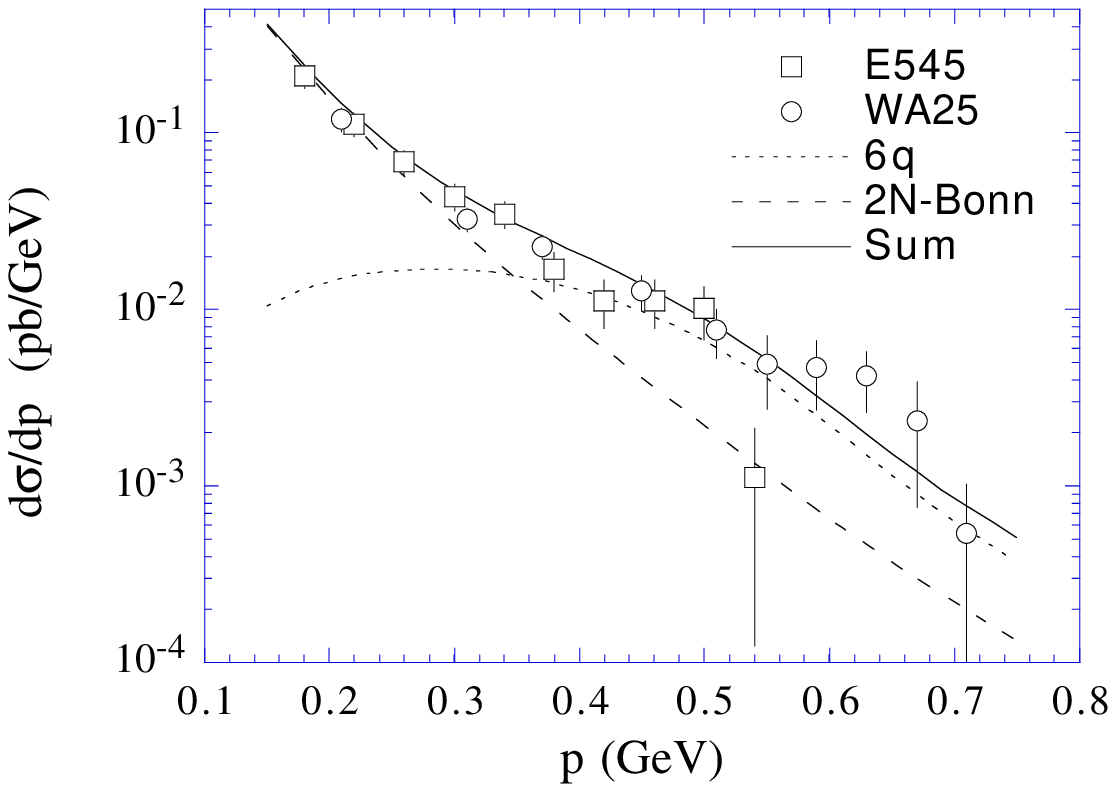}


\epsfxsize 3.45in 
                  \epsfbox{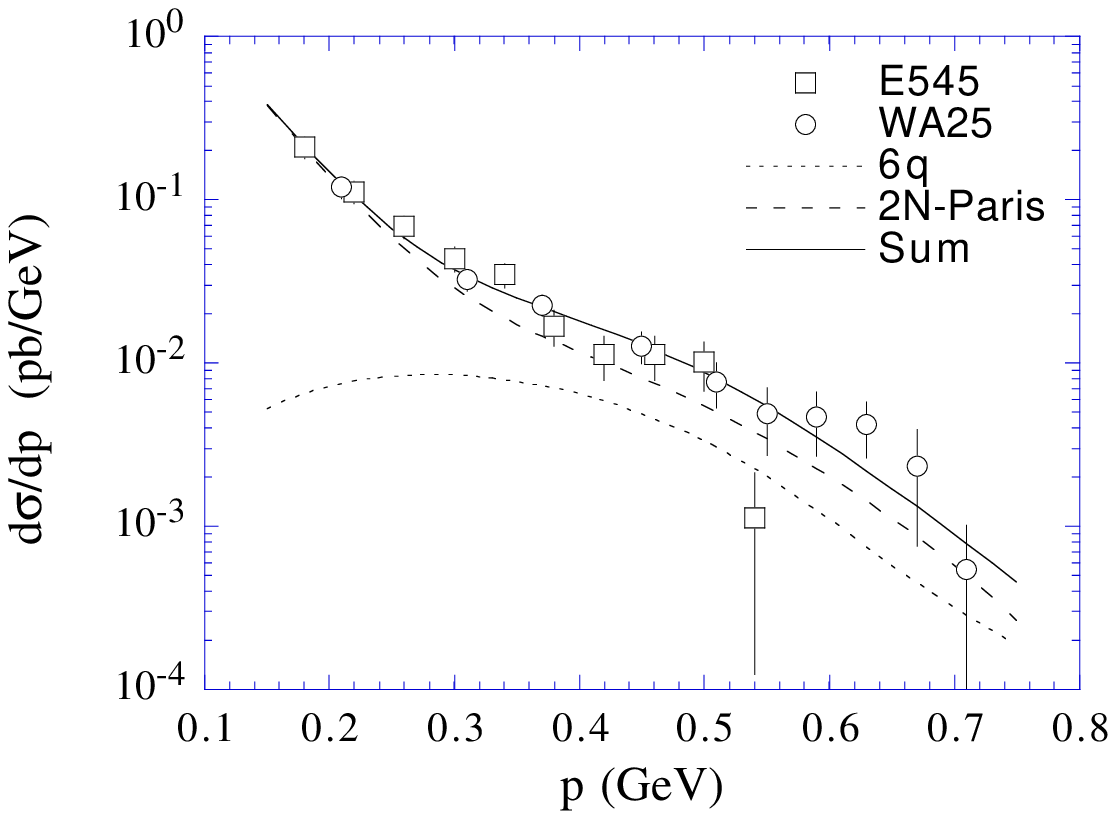}

\caption{Comparison of the backward proton spectrum in $\nu + d
\rightarrow \ell^- + p_B + X$ to calculations using the Bonn
(energy independent OBEPQ) wave function and the Paris wave function,
each using a $z^5(1-z)^3$ form in the fragmentation function and
with a 2\% 6q contribution for the Bonn wave function case, 1\%
for the Paris case.  As in Fig.~\protect\ref{han53}, similar results
follow with a $z^3(1-z)^3$ form and a 4\% or 2\% 6q contribution
for the respective cases.}
\label{all53paris}

\end{figure}

Fig.~\ref{all53paris}a shows what one of the Bonn wave functions, the
energy independent OBEPQ~\cite{bonn87} produces for the backward proton
spectrum.  Though a more complete and realistic wave function, the
results are not strikingly different than those from the Hulth\'en
wave function.  Fig.~\ref{all53paris}b shows a corresponding plot for
the Paris wave function.  All these potential models of the
nucleon-nucleon interaction give representations of our new data which 
are well below the data for the high momentum backward 
protons.  At a momentum of 500 MeV (c = 1),  a considerable contribution
must be added to each to bring them close to 
the data being presented: For the Hulth\'en, as noted above, the
additional amount needed is 800\%;  for the Bonn model an increase of
350\% is needed (see Fig.~\ref{all53paris}a) and for ths Paris model, an
additional 62\% is needed to bring theory and experiment into agreement.
The logarithmic scale for $d\sigma / dp$ in these figures 
misleads the eye.  But, in  Fig.~\ref{all53paris}b, it is 
clear that the Paris curve is below the majority of the data error bars.
To make a more compelling statement, we give a simple statistical 
comparison of the Paris curves with our data in the momentum
range, $0.2 < p < 0.5$ GeV.  The squares of the differences between the
center of the data points and the dashed Paris and solid Paris + 6q
curves divided by the error bar is calculated.  The result is a value of
$\chi^2$ per data point of 0.8 for the solid curve and and 3 for the
dashed curve, corresponding to confidence levels of 0.5 and 0.003, 
respectively, where the Particle Data Groups, graphs are used as a most
universal particle physics convention.

\section {Discussion and Conclusions}

We have seen that a small amount of 6q cluster in a deuteron can
explain the backward proton data.  It is still possible that some 2N
wave function with an increased amount of probability at high momenta
could also explain the data.  Therefore it is of interest to find
other signatures that could signal the presence of the 2N or 6q
states.  We will mention two possibilities, one if a polarized
deuteron target is available, another if there is enough data to
bin in both $x$ and $p$, and then conclude.

\begin{figure}

\epsfxsize 3.3in \epsfbox{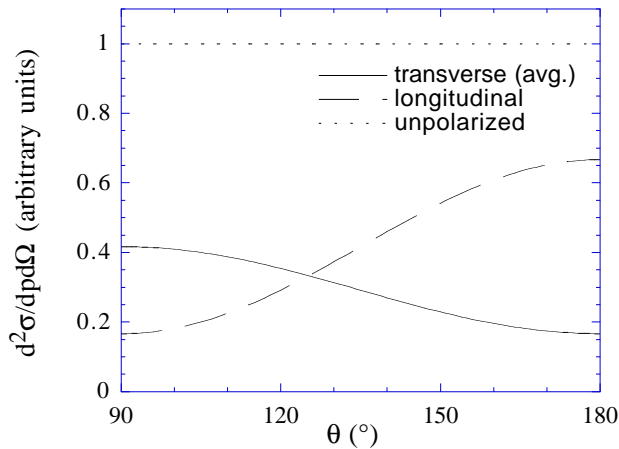}
\caption{Angular distribution of backward protons assuming a
polarized deuteron, the 2N model, and a momentum at which the
D-state dominates. The solid curve is for the average over the two
transverse polarizations of the deuteron, the dashed curve is for
the longitudinal polarization, and the dotted line shows the sum
over all polarizations.}
\label{angulardistribution}

\end{figure}

If a polarized target is available, then the 2N model leads to
characteristic variations of the backward  proton angular
distribution.  At low momentum, the wave function is mostly
S-wave and the backward distribution is angle independent
regardless of the deuteron's polarization.  At a momentum where
the D-state dominates (about 400 MeV for the Paris wave function,
as in Fig.~\ref{compare}), a polarized deuteron has a
non-isotropic spatial wave function. 
Fig.~\ref{angulardistribution} shows the angular distribution of
backward protons from the D-state for both a longitudinally
polarized deuteron (i.e., $J_z = 0$ with quantization direction
along the incoming current) and for the average of the two
transverse polarizations.  (The latter would be for either
transverse polarization in the analogous electromagnetic case.) 

If we can bin data in both $x$ and $p$, then we can define a
``two-nucleon test ratio.'' This is simply the ratio of the
observed differential cross section for backward proton scattering
to the cross section for scattering off a neutron with an
appropriate momentum shift.  The latter is intended to be just what
would be expected for the 2N model, with the wave function factor
removed.  Explicitly,
\begin{equation}
{R_1} = {\sigma_{meas}(x,y,\alpha,p_T)\over
         \sigma_{nX}(x,y,\alpha,p_T)}  ,
\end{equation}

\noindent where the denominator is
\begin{eqnarray}
\sigma_{nX} &=& {d\sigma \over 
             dx\, dy}(\nu n \rightarrow \mu^- X) \nonumber \\
    &=& K (2-\alpha)
     \left[ D_n(\xi)+S_n(\xi)+(1-y)^2 \bar U_n(\xi) \right].
\end{eqnarray}

\noindent If the 2N model is correct, then
\begin{equation}
R_1 = |\psi(\alpha,p_T)|^2 .
\end{equation}

Thus, we can test for the 2N model by plotting $R_1$ vs. $x$ at fixed
$\alpha$ and $p_T$, or at fixed $p$.  If the 2N model is right,
such a plot would produce just a simple horizontal line.  The
crucial question is how different a result a 6q cluster would
give.  We have elaborated on this question in Ref.~\cite{cl95} for
the electromagnetic case.  If the 6q cluster dominates at some
fixed backward proton momentum, it gives a curve for $R_1$ vs. $x$
that varies by a factor of roughly two from peak to valley. It
should be easily distinguishable from the 2N expectation.

In conclusion, we have studied the production of backward protons in
neutrino- and antineutrino-deuteron scattering, and compared the
existing data to one model.  The backward proton data from the E545
Fermilab experiment shown in this paper has not previously been
published, although it has appeared in talks~\cite{kaf83}.  The
model we have considered is an incoherent sum of contributions from
2N and 6q components of the deuteron.  None of the 2N models that we
have looked at has by itself a large enough high momentum tail to
explain the backward proton data above about 300 MeV/$c$.  If we add
a 6q component, we get straightforwardly a good match to the shape
of the backward proton spectrum at high momentum.  If the
probability of the 6q cluster is one to a few percent in the
deuteron, then the 6q contribution accounts well for the observed
normalization of the data at high momentum, while adding negligibly
to the 2N contribution below about 250 MeV/$c$~\cite{italians}.

We consider this a good indication that 6q configurations exist in
the deuteron and can be observed in certain circumstances.  It is
however not ironclad proof since in principle it may be possible that
some 2N wave function with an enhanced high momentum tail could also
explain all the backward proton data.  But one should realize that a
wave function gotten from a potential that is fit to data, including
low energy nucleon-nucleon scattering data, is matching a Nature that
may contain 6q cluster effects and must mock them up somehow in the
context of its own degrees of freedom.  This means that a good fit to
the data with just a 2N wave function is not in its own turn
ironclad proof against a 6q cluster.  Hence we have added suggestions
of further tests that may eventually argue directly against the 2N
models.

\section*{Acknowledgments}

We express our appreciation to the Fermilab E545 collaboration
for providing their unpublished $\nu d$ backward proton spectrum. 
CEC thanks the NSF for support under Grant PHY-9600415,  and also
O. Benhar,  S. Liuti, and V. Nikolaev for useful comments.  CEC
and KEL both thank Fermilab for its hospitality while part of this work
was done.

\end{document}